# Evaluating ECG Capturing Using Sound-Card of PC/Laptop

B. N. Patel[1], D. N. Shah[2]

[1,2]Department of instrumentation and control, Sarvajanik College of Engineering and Technology Surat, Gujarat, INDIA

*Abstract*

*The purpose of the Evaluating ECG capturing using sound-card of PC/Laptop is provided portable and low cost ECG monitoring system using laptop and mobile phones. There is no need to interface micro-controller or any other device to transmit ECG data. This research is based on hardware design, implementation, signal capturing and Evaluation of an ECG processing and analyzing system which attend the physicians in heart disease diagnosis. Some important modification is given in design part to avoid all definitive ECG instrument problems faced in previous designs. Moreover, attenuate power frequency noise and noise that produces from patient's body have required additional developments. The hardware design has basically three units: transduction and conditioning Unit, interfacing unit and data processing unit. The most focusing factor is the ECG signal/data transmits in laptop/PC via microphone pin. The live simulation is possible using SOUNDSCOPE software in PC/Laptop. The software program that is written in MATLAB and LAB-View performs data acquisition (record, stored, filtration) and several tasks such as QRS detection, calculate heart rate*.

## 1. Introduction

**What is an ECG?**

The Heart is a muscle formed that allows it to act as a pump for blood circulation in a body. The heart pumps blood when the muscle cells making up the heart wall contract, generating their action potential. This potential creates electrical currents that spread from the heart throughout a body.The electrical potential between various locations shows differences due to spreading electrical currents in the body. The electrical potential can be detected and recorded by placement of electrodes on skin. The cardiac electrical potential waveform generated by these bio-potentials is called the Electrocardiogram (ECG).

**Electro-conduction system of Heart**

The conduction system of the heart is shown in figure 1. It consists of the Senatorial (SA) node, bundle of his, Atrioventricular (AV) node, the bundle branches, and Purkinje fibres. The SA node serves as a pacemaker for the heart and it provides the trigger signal. It is a small bundle of cells located on the rear wall of the right atrium, just below the point where the superior vena cava is attached. The SA node fires electrical impulses through the bioelectric mechanism. It is capable of self excitation but is under control of the CNS so that the heart rate can be adjusted automatically to meet varying requirements.

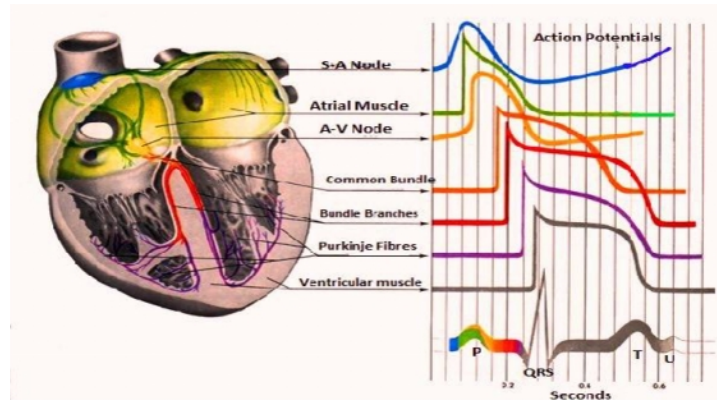

Fig 1: Electro-Conduction System of Heart

When the SA node discharges a pulse, then electrical current spreads across the atria, causing them to contract. Blood in the atria is forced by the contraction through the valves to the ventricles. The velocity of propagation for the SA node action potential is about 30cm/s in the atrial tissue. There is a band of specialized tissue between the SA node and AV node, however in which the velocity of propagation is faster than it is in atrial tissue. This internal conduction pathway carries the signal to the ventricals. The muscle cells of ventricles are actually excited by purkinje fibres. The action potential travels along these fibres at the much faster rate on the order of 2 to 4 m/s. The fibres are arranged in two bundles, one branch on left and another is on right.

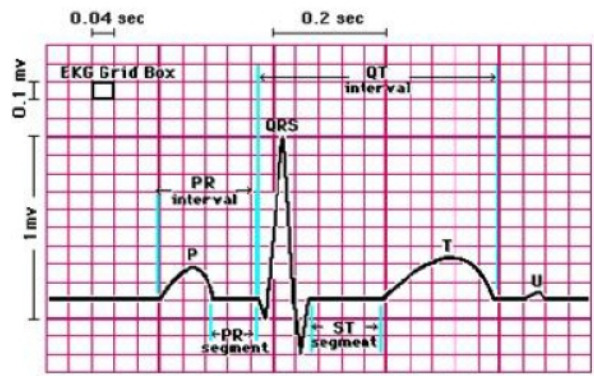

Fig 2: Typical ECG Waveform

Conduction in the fibres is very rapid. The action potentials generated in SA node stimulates the muscle fibres of the myocardium, causing them to contract. When the muscle in the contraction, it is become shorter, and the volume of the ventricular chamber is less, so blood is constricted out. The shortening or tensing of so many muscles at one time creates a mass electrical signal that can be detected by electrodes place on the surface of the patient's chest. This electrical dispatch can be plotted as a function of time and the resultant waveform is known as Electrocardiogram (ECG).

### Different Methods of ECG Capturing

There are various Methods available of ECG capturing and evaluating.

1) ECG Machine
2) Portable ECG Monitor
3) ECG Readout Device
4) PC based ECG system
5) Using sound card of PC

## 2. Circuit Design

### Introductory Description of Circuit Design

The aim is to design a sensitive amplifier circuit that can detect ECG (Electrocardiogram) signals found from electrodes applied at the left arm (LA), Right arm (RA) and right leg (RL). The difference between the right-arm (RA) lead and the left-arm (LA) lead is amplified by circuit with the right-leg (RL) lead as the ground or reference node. The circuit needs a two stage instrumentation amplifier which has buffering stage and amplification stage. After that filtering stage is required for signal filtration.

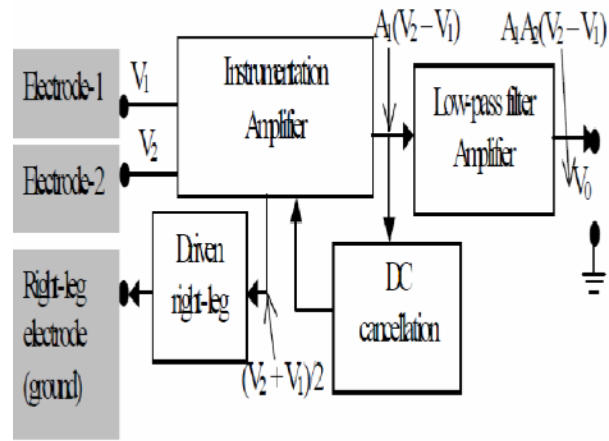

Fig 3: General Block Diagram of ECG Amplifier

A general block diagram of a modern ECG amplifier is shown in Fig. 3. It is designed around an instrumentation amplifier. In addition, the instrumentation amplifier has feature of an automatic offset (DC) cancellation circuit to keep the output always zero averaged. The electrical safety is provided by driven-right leg circuit and the interference is reduced under normal operational conditions. Low-pass filter provides further amplification and limit the bandwidth of the circuit.

**Electrode placement**

ECG is recording most commonly between the Right Arm (RA) and the Left Arm (LA). Sometimes another two combinations using the Left Leg (LL) are also used clinically (RA–LL and LA–LL). For common ground of the Instrumentation amplifier, there are one another electrode connects to the patient. This is attached to the right leg.

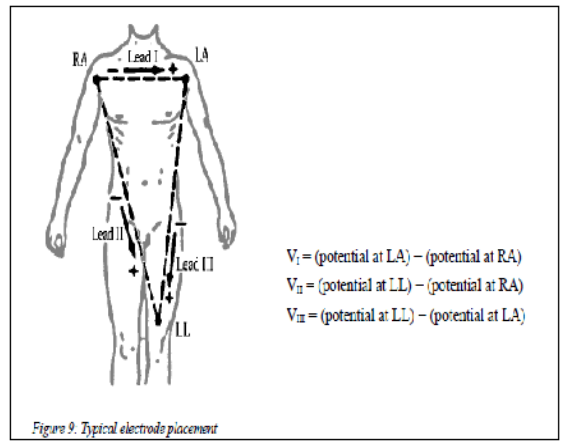

Fig 4: Typical Electrode Placement

The movement of electrode with respect to the electrolyte is mechanically disturbed. So, the distribution of charge at the interface and results in a momentaneous change of the half–cell potential until equilibrium can be restored. If one electrode is moved while the other remains stable, a potential difference appears between the two electrodes during this movement. Due to this kind of movement the potential is referred to as movementartifactand this can be a grievous cause of interference in the measurement of ECG.

**Theoretical Design**

To fulfil the requirements of our ECG amplifier, we need to design a cascade circuit, which is a combination of a Instrumentation Amplifier, a Low Pass Filter, a High Pass Filter and a gain stage. For reducing noises, the order of cascade stages is considered. For example, in the following cascade Figure, the output noise is

$$((n_1 * A_1 + n_2) * A_2 + n_3) * A_3 = A_1 A_2 A_3 * n_1 + A_2 A_3 * n_2 + A_3 * n_3, \; A_1 > A_2 > A_3.$$

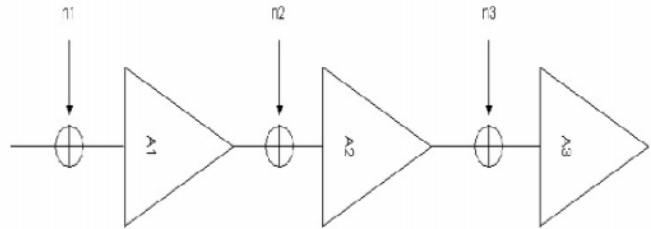

Fig 5: Cascade Design of Amplifier

Cascade design is on the basis of placing high gain stages in the signal path. All the same, the High Pass Filter stage should be placed immediately after the differential amplifier to remove DC offset.

## Instrumentation Amplifier AD620

The AD620 is a low cost, high accuracy of 40 ppm with maximum nonlinearity instrumentation amplifier that requires only one external resistor to set gains of 1 to 1000. The AD620 has low offset voltage of 50 mV max and offset drift of 0.6 mV/°C max.

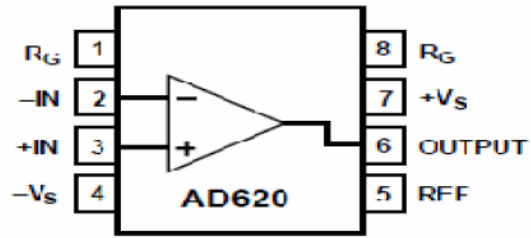

Fig 6: Pin Diagram of AD620

Furthermore, it has low noise, low input bias current and low power which makes it well suited for medical applications such as ECG and non-invasive blood pressure monitors. Gain of the AD620 is set by connecting a single external resistor RG, as shown commonly used gains and RG resistor values. The internal gain resistors R1 and R2. Trimmed to an absolute value of 24.7K , allowing the gain to be programmed accurately with a single external resistor. These resistors are affected by the accuracy and temperature coefficient included in the gain accuracy.

$$G = 1 + (49.4 \text{ k}/RG) \ldots\ldots\ldots (1)$$

| 1% Std Table Value of $R_G$, Ω | Calculated Gain | 0.1% Std Table Value of $R_G$, Ω | Calculated Gain |
| --- | --- | --- | --- |
| 49.9 k | 1.990 | 49.3 k | 2.002 |
| 12.4 k | 4.984 | 12.4 k | 4.984 |
| 5.49 k | 9.998 | 5.49 k | 9.998 |
| 2.61 k | 19.93 | 2.61 k | 19.93 |
| 1.00 k | 50.40 | 1.01 k | 49.91 |
| 499 | 100.0 | 499 | 100.0 |
| 249 | 199.4 | 249 | 199.4 |
| 100 | 495.0 | 98.8 | 501.0 |
| 49.9 | 991.0 | 49.3 | 1,003 |

Table 1: Values of Rg for settling Gain of AD620

The stability and temperature drift of the external gain setting resistor, RG, also affects gain. RG's contribution to gain accuracy and drift can be directly inferred from the gain equation. There are many important features of AD620 such as gain set with one resistor, wide range power supply, 100dB min CMRR (common-mode rejection ratio), low noise and excellent DC performance. AD620 is used in many applications like portable battery operated system, physiological amplifier: EEG, ECG, EMG, multi channel data acquisition, ECG and medical instrumentation etc.

**Designed Circuit**

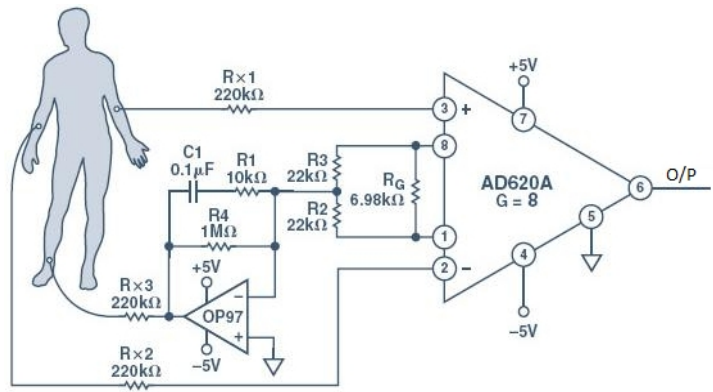

Fig 7: ECG Circuit with Right Leg Driven Circuit

**Frequency Adjustment**

The average heart rate of a person is around 1.1Hz and the signal level is very weak thus to avoid interference from other signals, a band pass filter is needed. The desired frequencies (between 0.05 Hz and 150Hz) can be achieved via operational amplifier, capacitors and Resistors where the values are found for the high pass and low pass respectively.

Low pass   $0.05 Hz = \frac{1}{2\pi RC}$

High Pass   $150 Hz = \frac{1}{2\pi RC}$

For High Pass filter        R = 3.18 M   , C= 1μF

For Low Pass filter        R = 106 K   , C= 0.01μF

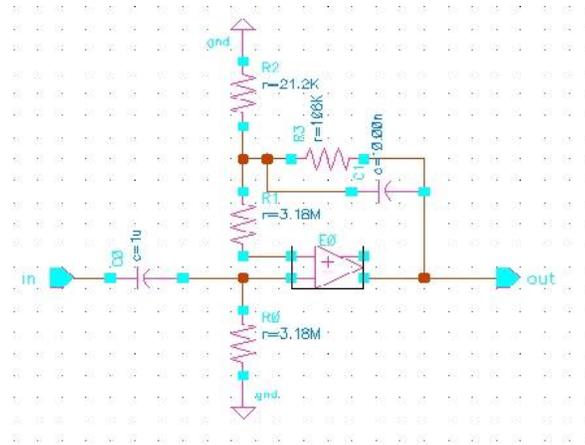

Fig 8: Filter Circuit for ECG

Thus the filter supplies a gain of 5. The Resistor 3.18M   connected to non-inverting input is for keeping the balance and symmetry.

## Right Leg Driven Circuit

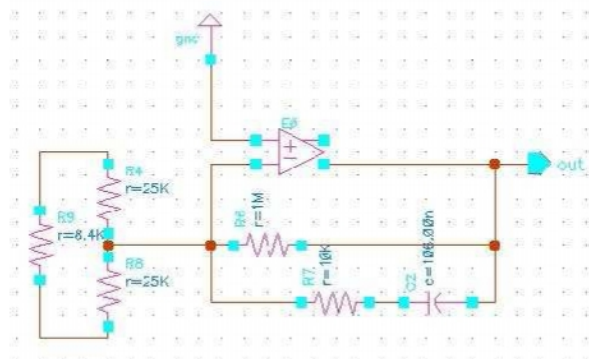

Fig 9: Right Leg Driven Circuit

The aim of right leg driven circuit is reducing the effect of noise. The common mode signal taken from the both ends of the Rg (gain resistance of AD620) is given back to body as a reference.

$$G = \frac{1M\Omega}{\frac{25K\Omega}{25K\Omega}}$$

The values of the resistances are 25KΩ for the input resistance. Also a low pass filter with 150Hz cut-off frequency is existent with R = 10KΩ and c =106nf. The gain of the circuitry is defined by 1MΩ resistance.

## Output DC level Shifting

The last stage of the circuit is both designed to provide the necessary gain and the necessary voltage shifting to make the signal appear on the limits is 0 to 5 volts. This stage is expected to have a gain of 25. Since the amount of voltage shifting is unknown the voltage shifting is applied after the first ECG signal is observed. Initially an inverting amplifier with R1=5KΩ and R2=125KΩ used.

## Realization of The circuit

The circuit is first realized on breadboard. Some of the values have changed due to the availability of the resistors and capacitors. For the operational amplifier op07 is used in all stages. While the circuitry is being built each stage is checked by a given signal which would not cause clipping.

## Changes with Realization

1) AD620

Installing Rg as 8.4KΩ gave unwanted gain so the value had to be increased to decrease the gain. Inserting a 27KΩ resulted with a gain of 10, even though it was unexpected, in order to not using a larger resistance, the 27KΩ is accepted and the extra gain is corrected by changing the last stage.

2) Band-pass stage

For the high-pass filter a resistance of 3.18MΩ is used. The resistance is in fact labelled as 3MΩ, however its exact value was 3.18MΩ. For the low-pass filter the 106 is realize as an 112KΩ and the 21.2KΩ is realize as a 21.8KΩ (series combination of 3.8KΩ and 18KΩ)

3) Output DC Level Shifting

The amount of shifting is done by supplying is done by supplying the inverting input and supplying the shifting dc voltage from the non-inverting input, thus the amount of dc input can be found by dividing the desired DC voltage level by non-inverting gain of the stage.

Fig 10: Gain Stage and DC level Shifter

However due to non-ideal of op07 the non-inverting input lowered the gain. With an interactive method the correct values are found. The input dc voltage is supplied from Vdc by a voltage divider.

## Modification

To prevent the signal being lost in the 50Hz interface, a 50Hz notch filter is designed and implemented.

Fig 11: Schematic of Notch Filter

## Complete ECG Circuit

Fig 12: Complete ECG Circuit

## Soundcard of Laptop

A sound card also known as an audio card which has facilitates the input and output of audio signals to and from a computer under control of computer programs. Typical uses of sound cards include providing the audio component for multimedia applications such as music composition, editing video or audio, presentation, education, and entertainment (games). Many computers have in built sound capabilities, while others require further soundcard expansion cards to provide for audio potentiality.

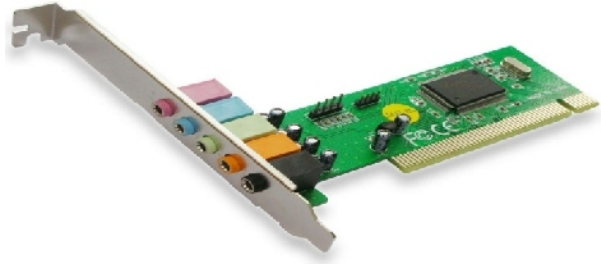

Fig 13: Soundcard Hardware

Sound card has usually functioned of analogue-to-digital converter (ADC), which converts recorded or generated analogue data into a digital format. The output signal is connected to an amplifier, headphones, or external device using standard interconnects. For higher data rates and multiple functions, there is more advanced card commonly include more than one chip.

## Microphone Pin Configuration

The diagram below illustrates how to configure a standard stereo microphone plug. The tip of the pin is the left channel, the ring type metal portion is the right channel, and the rest of the pin is the ground. There is one plastic ring between two channels which separates the channel and ground. Use a multimeter or continuity tester to determine the channel identifications of the solder logs.

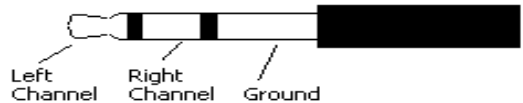

Fig 14: Microphone Pin

1 – Signal (audio) out left channel/Channel-1
2 – Signal (audio) out right channel/Channel-2
3 – Ground – common for microphone and audio out.

## Heart Pulse Detector Circuit

Photoplethysmography (PPG) is a non-invasive method of measuring the variation in blood volume in tissues using a light source and a detector which is the principle of heart pulse detector circuit.

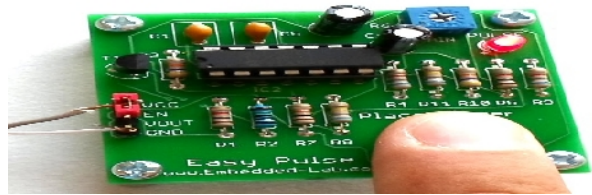

Fig 15: Heart Pulse Sensor

Since the change in blood volume is synchronous to the heart beat, this technique can be used to calculate the heart rate. For reflectance PPG, the light source and the light detector are both placed on the same side of a body part. The light is emitted into the tissue and the reflected light is measured by the detector. PPG can be applied to any parts of human body. In either case, the detected light reflected from or transmitted through the body part will fluctuate according to pulse rate blood flow caused by the beating of the heart.

## 3. Software Description

### Sound Card Scope

The PC based Soundcard Oscilloscope obtains its data with 96 kHz and 16 Bit resolution from the soundcard. The data source can be selected in the Windows mixer such as Microphone, Line-In or Wave.

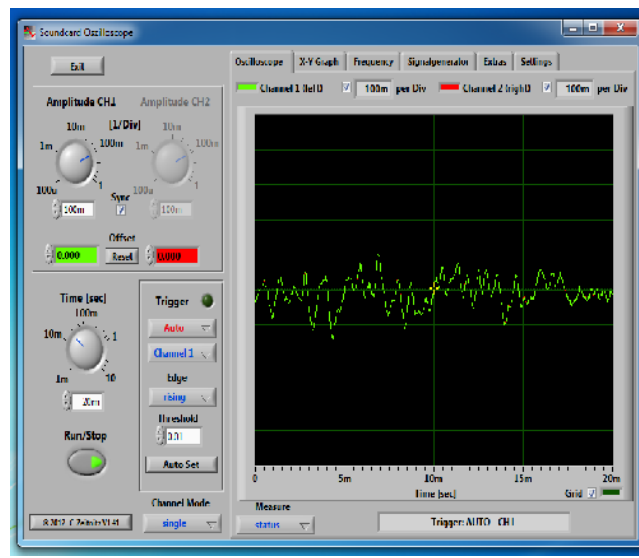

Fig 16: Sound Scope Display

The frequency range depends on which kind of a soundcard already in laptop/pc, but 20-20000Hz should be possible with all modern cards. The oscilloscope contains in further a signal generator for 2 channels for Sine, Square, Triangular and Saw tooth wave forms in the frequency range from 0 to 20 kHz. These signals are available at the speaker output of the sound card.

## Implementation Steps in Sound Scope

1) Run sound scope software.
2) Go settings and Check windows Sound parameters

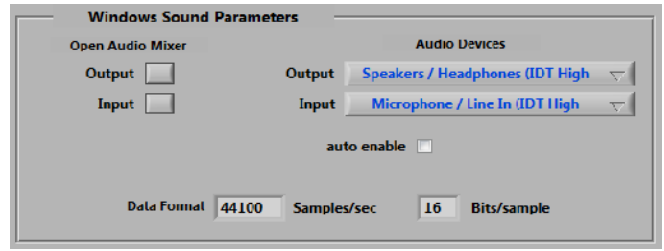

3) Change Input tab as Microphone / Line In.
4) Select the channel-1 / Channel-2 as microphone pin wire connection.
5) Set the Amplitude and Time Division

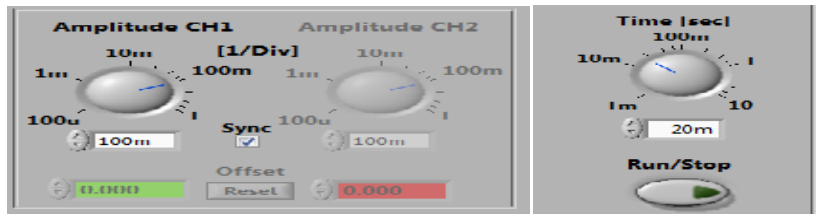

6) Run the simulation and Check the waveform.
7) Pause simulation and store data file.

## MATLAB Simulink

### Define soundcard in MATLAB

For data acquisition of ECG signal, it is very important to interface sound card in MATLAB. There is a facility available to interface sound card in MATLAB for particular frequency up to 96 kHz. It records data fro given time interval after that we have to plot data and apply different operation related evolution and analysis of ECG signal.

MATLAB code for Sound card interface

```
AI = analoginput('winsound');
addchannel(AI, 1);
Fs = 5000;% Sample Rate is 100 Hz
set (AI, 'SampleRate', Fs)
duration = 5;       % 5 second acquisition
set(AI, 'SamplesPerTrigger', duration*Fs);
start(AI);
data = getdata(AI);
```

International Journal of Instrumentation and Control Systems (IJICS) Vol.4, No.1, January 2014```
delete(AI)
plot(data,'DisplayName','data','YdataSource','data');
figure(gcf)
```

**Filter of signal in MATLAB**

Some common ECG Filtering tasks are performed which are Baseline wander filter and Power line interference filter. When filtering any biomedical signal care should be taken not to falsify the desired information in any condition. A major intrest is how the QRS complex affects the output of the filter. Sometimes the filter often carry out a large unwanted impulse and possible distortion caused by the filter should be cautiously quantified.

**Apply Butterworth Filter /Base line Filter**

Baseline filter is included extragenoeous low-frequency for high-bandwidth components and it can be caused by movements of electordes which effects electrode impedance, Respiration and Body movements. Which can cause troubles to analysis mostly when analise the low-frequency ST-T segment.

MATLAB Code :

```
sig = data
plot(sig)
[A,B] = BUTTER(2,0.002)
subplot(211),plot(data);
subplot(212),plot(filter(A,B,data));
```

**Apply Notch filter / Power Line Filter**

Electromagnetic fields from power lines can cause 50/60 Hz sinusoidal disturbance, possibly companioned by some of its constant frequency**.** Such noise can cause problems interpreting low-amplitude waveforms and unauthentic waveforms can be introduced. Naturally care should be taken to keep power lines as far as possible and shield and ground them but this is not always possible everywhere.

MATLAB Code :

```
input=data
plot(input)
fs=750;
N=length(input);
t = ((0:length(input)-1)./fs); %Time Interval
t=t';
figure(1)
plot(t, input)
xlabel('Sampling frequency')
ylabel('Volts')
```

23



```
Ys = fft(input)/N;
i_space=linspace(0,.5,N/2); %adjusting the frequency range at particular interval
frq = fs*i_space;
Ys = Ys(1:ceil(N)/2);
plot(frq,2*abs(Ys))
xlabel('Frequency in Hz')
ylabel('|Y(f)|')

%intialization

F_0=0.33;
Delta_F=0.1;
[b,a] = iirnotch (F_0,Delta_F);%creates coefficient vectors a and b
fvtool(b,a);
refined=filter(b,a,input); %filter signal
[H Hf] = freqz(b,a,N); %returns transfer function
amp=abs(H);
figure(2)
plot(Hf*fs,amp)
xlabel('Frequency in Hz')
ylabel('|H(f)|')

 %Attempting different values of delta F

subplot(3,1,1)
plot(t,refined)
title('Refined ECG Signal when delta F=1')
xlabel('Sampling Frequency')
ylabel('volt')
F_0=0.6;
Delta_F=0.001;
[b,a] = iirnotch (F_0,Delta_F);%creates coefficient vectors a and b
refined=filter(b,a,input); %filter signal
subplot(3,1,2)
plot(t,refined)
title('Refined ECG Signal when delta F=5')
xlabel('Sampling Frequency')
ylabel('volt')

 F_0=0.9;
Delta_F=0.0001;
[b,a] = iirnotch (F_0,Delta_F,fs);%creates coefficient vectors a and b
refined=filter(b,a,input); %filter signal
subplot(3,1,3)
plot(t,refined)
title('Refined ECG Signal when delta F=10')
xlabel('Sampling Frequency')
ylabel('volt')
```



International Journal of Instrumentation and Control Systems (IJICS) Vol.4, No.1, January 2014

```
%interference
interference=input-refined;
plot(t, interference)
title('Interference')
xlabel('Sampling Frequency')
ylabel('volt')
```

**Heart Beat Count**

The data signal hold and sample with particular sample frequency for counting the heart pulses. Then count the dominant peak of signal which corresponds to heart pulses. Peaks are defined to be samples greater than their two nearest neighbours and greater than 1 and divides beat counted by signal.

MATLAB Code:
```
%progaram to determine the BPM of ECG signal

sig = data;
plot(sig)
xlabel('Samples');
ylabel('Electrical Activity');
title('Ecg Signal sampled at 100Hz');
hold on
plot(sig,'ro')

 %count the dominant peaks in the signal(these 25orrespond to heart beats)
% - peaks are defined to be samples greater than their two nearest neighbours and greater than 1

beat_count = 1;
for k = 2 : length(sig) – 1
if(sig(k) > sig(k-1) & sig(k) > sig(k+1) & sig(k) > 1) %k
%disp('Prominent Peak found')
beat_count = beat_count + 1
end
end

%Devide the beats counted by the signal duration (in minutes)
beat_count = 1;
fs = 4000;
N = length(sig);
duration_in_seconds = N*0.1/fs;
duration_in_minutes = duration_in_seconds/60;
BPM = beat_count/duration_in_minutes

sig = sig(1:500);
hold off
plot(sig)
```





## LAB-View

This technical description will introduce the concepts required to build a basic system with LAB-View. The most important building blocks for any LAB-view application, including the front panel, block diagram, palettes, controls, and indicators. Graphical Programming Basics see how to connect functions and work with a variety of data types when constructing applications. Common Tools view a collection of important tools and common user functions that all users should be familiar with LAB-View.

### Lab View Design To get ECG data

The basic block diagram of the ECG signal condition is shown in fig 17. The filtrations procedure describes below and detail of specification of software components.

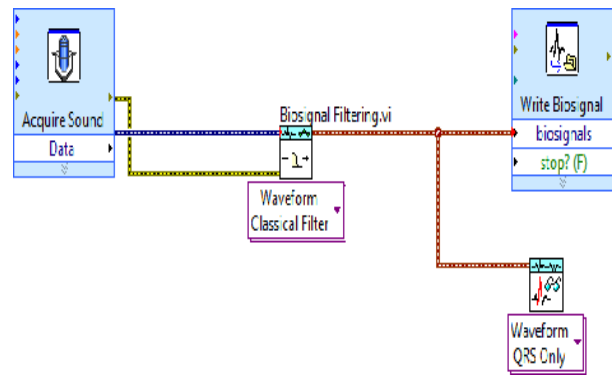

Fig 17: Block Diagram of ECG in LAB-View

### Acquire Sound Block

**[1] Error in** describes error conditions that occur before this function runs and this input provides standard error in functionality.
**[2]Path** out identifies the ECG data stored in wave file.

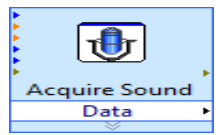

**[3] Total number of samples** have total number of channels selection and the number of bits per sample in the wave file.
**[4]Format** shows the sample rate, the number of channels and the number of bits per sample in the wave file.
**[5]Sample rate (S/s)** selects sampling rate for the wave file. Common rates are 44,100 S/s,22,050 S/s and 11.025 S/s.
**[6]Number of channels** specifies which channel of soundcard in the wave file. This input can accept as many channels as sound card supports. For most sound cards 1 is Mono and 2 is Stereo.
**[7]Bits per sample** are the quality of each sample in bits. Common resolutions are 16 bits and 8 bits but 16 bits are more accurate than 8 bits.





**[8]Error out** contains error information. This output provides standard error out functionality.

**Biosignal Filtering (Waveform Classical / Array Filter) Block**

**[1]Type** specifies the classical filter to apply to input biosignal 0 Low-pass,1 High-pass and 2 Band-pass

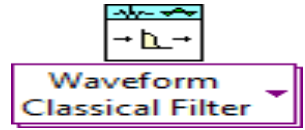

**[2]Freq specs** specify the frequency specifications for the classical filter.
**[3]Fpass 1** specifies the first pass band edge frequency in hertz. The default is 0.2.
**[4]Fpass 2** specifies the second pass band edge frequency in hertz. This VI ignores Fpass 2 for low pass and high pass filters. The default is 0.
**[5]Fstop 1** specifies the second stop band edge frequency in hertz. The default is 0.3.
**[6]Fstop 2** specifies the second stop band edge frequency in hertz. This VI ignores for low pass and high pass filters. The default is 0.1.
**[7]Ripple specs** specifies the ripple specifications for the classical filter.
**[8]Pass band** specifies the ripple level in the pass band. The default is 0.1.
**[9]Stop band** specifies the ripple level in the stop band. The default is 60.
**[10]Error in** describes error conditions that occur before this function runs.
**[11]Out** returns the filter that this VI applies to the input biosignal.
**[12]Error out** contains error information. This output provides standard error out functionality.

**ECG Feature Extractor (Waveform QRS) Block**

**[1]QRS detector parameter specifies** the parameters that this VI uses to detect QRS waves.
**[2]Rough highest heart rate** determines the approximate highest heart rate of the ECG signal in beats per minute (bpm) and the default value is 60 bpm.
**[3]Peak detection initial threshold** determines the initial threshold for QRS complex detection.

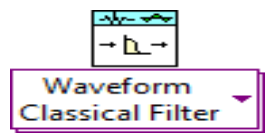

**[4]Threshold factor** defines the factor that VI uses to determine the threshold for separating noise peaks and QRS waves. Threshold factor must be greater than 0 and less than 1.
**[5]Error in** describes error conditions that occur before this function runs and this input provides standard error in functionality
**[6]QRS time** returns the occurring times of detected QRS complexes. This QRS time by using the input ECG signal, which might be a preprocessed signal.
**[7]QRS peak** returns the amplitudes of the detected QRS complexes. QRS peak through the input ECG signal, which might be preprocessed..
**[8]Error out** contains error information. This output provides standard error out functionality





## Bio-Medical Tool connectivity

The Biomedical Workbench in LAB-View Biomedical Toolkit provides applications for bio signal and biomedical image analysis. These applications enable to apply biomedical solutions using Lab VIEW. These applications to log and play biosignals, simulate and generate biosignals, analyzebiosignals, and view biomedical images. The application can acquire real world and real-time biomedical data by using biomedical sensors and National Instruments hardware. The applications in Biomedical Workbench is extract features from electrocardiogram (ECG) signals, to analyze heart rate variability (HRV), and to measure blood pressure.

## ECG Feature Extractor

ECG feature Extractor imports ECG signals from different file types. See Biosignal Viewer for file formats supported. It imports ECG signals from a data acquisition (DAQ) device and integrates robust

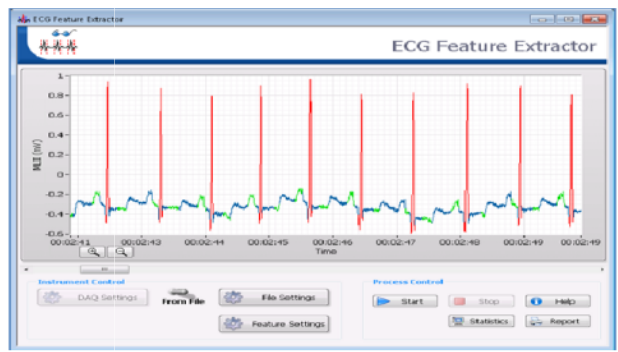

Fig 18: ECG feature Extractor

extraction algorithms to detect ECG features, such as the QRS Complex, P wave, and T wave. It transfers RR interval data to Heart Rate Variability Analysis application and exports ECG features reports for printing.

## Heart Rate Variability (HRV) Analyzer

HRV analyzer synchronizes RR intervals from the ECG Feature Extractor application and imports RR intervals from an electrocardiogram (ECG) file that the ECG Feature Extractor application generates or from a text file that contains RR intervals.





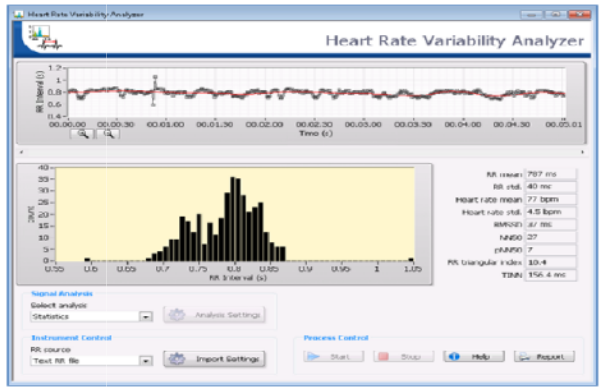

Fig 20: Heart Rate Variability Analyzer

It provides a variety of analysis methods for HRV analysis including Statistics (including histogram), Poincare plot, FFT spectrum, AR spectrum, STFT spectrogram, Gabor spectrogram, wavelet coefficients, DFA plot and recurrence map.

## 4. Results

### [1] Designed circuit output

Waveform in sound scope

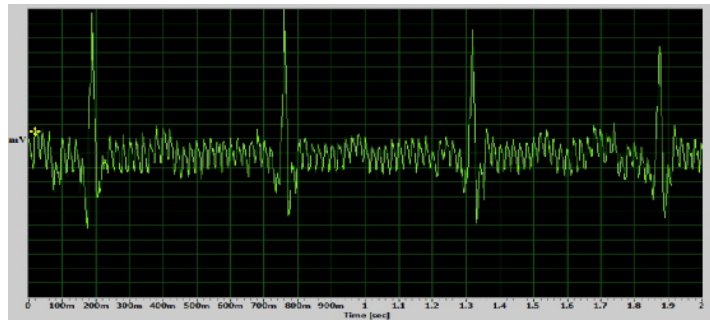

MATLAB Output

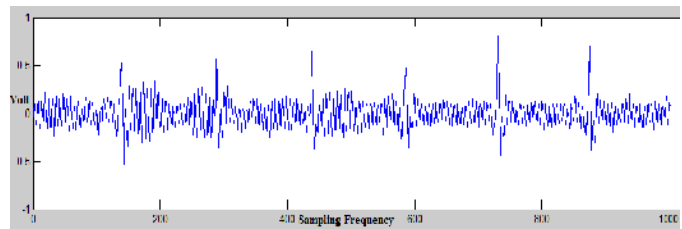



International Journal of Instrumentation and Control Systems (IJICS) Vol.4, No.1, January 2014

MATLAB Filtrations output

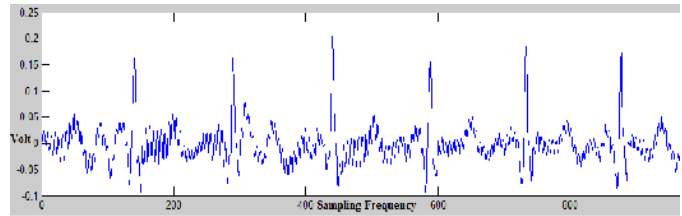

Heart Beat Result in LAB-View

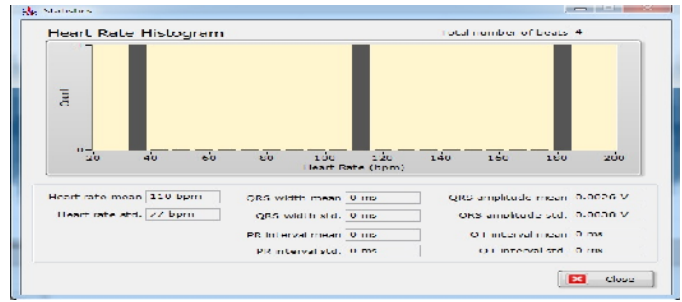

Filtrations in LAB-View

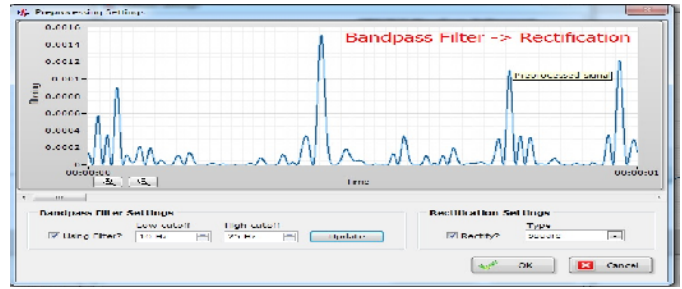

Heart Beat Result in MATLAB

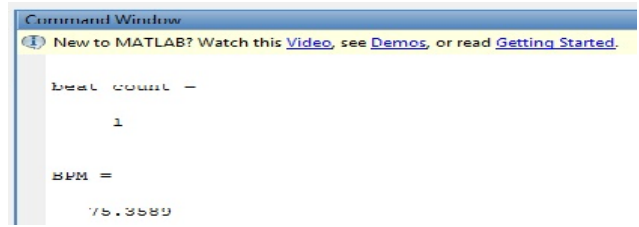

MATLAB Filtrations output

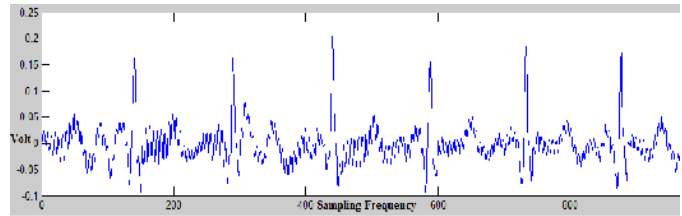

Heart Beat Result in LAB-View

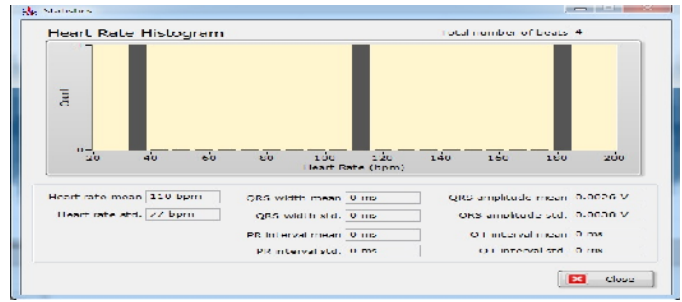

Filtrations in LAB-View

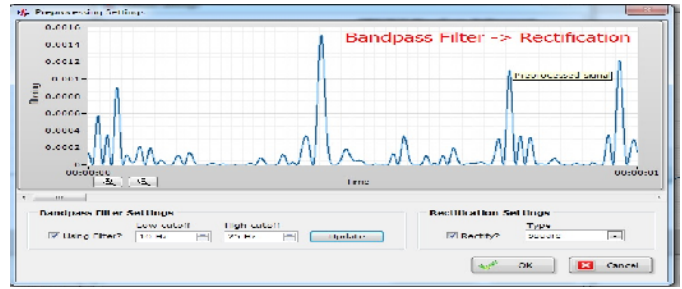

Heart Beat Result in MATLAB

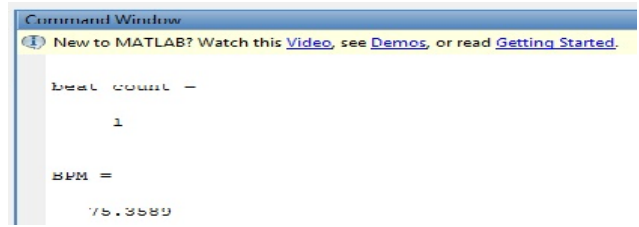



International Journal of Instrumentation and Control Systems (IJICS) Vol.4, No.1, January 2014

**[2] Heart pulse sensor output**

Digital Oscilloscope Output

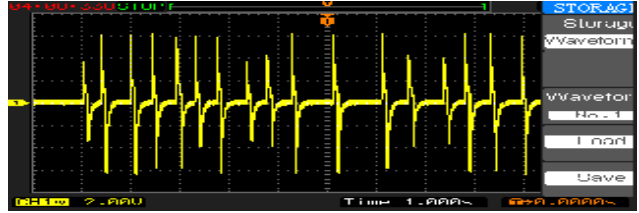

Sound Scope Output

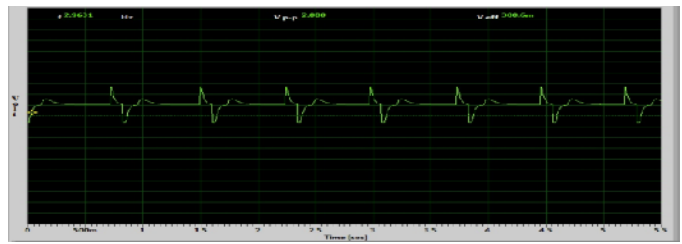

MATLAB Output

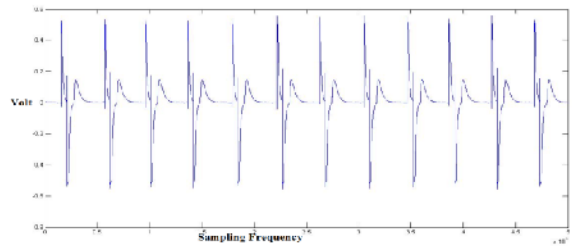

## 5. Conclusion

Evaluating ECG Capturing Using sound Card of PC/laptop is designed and performed without interfacing microcontroller or any other data transmission devices. The transformations of ECG signal from circuit to laptop via micro-phone jack which is already connected with inbuilt soundcard. This method has low cost and less complexity. Only it required basic knowledge of circuit design and programming skills. In circuit design, various electronics and bio-medical instrumentation's principle are used. AD620 is a specific instrumentation amplifier. This is used specially in bio-medical application because it has high CMRR and low input noise amplifier. There are high-pass, low-pass, band-pass and notch filters are designed for reduced frequency oriented noise.

The Sound-scope is a one of the software which shows the live or real time simulation of ECG signal which transmit through micro-phone pin. MATLAB is stored the data in sampling





frequency of soundcard. In which sample and hold process carry out and various operations such are filtration and beat count performed by programme. In LAB-View, the basic block diagram designed and each block has facility of set all function and variable value with bio-medical tool connectivity.

When the circuit is implemented, sensitivity of the circuit is very high. There is some minor change and movement affect the output wave. After, the take care by using special ECG lead cables and shield cable is used for interface a circuit. Circuit is totally cover with the aluminium foil paper so that, noise interference from environment and electrical & electronics components is reduced.

## Acknowledgement

We welcome this opportunity to express our devout appreciation and regards to our Head of Department Dr. UtpalPandya, Department of instrumentation & control Engineering, Sarvajanik college of engineering and Technology, Surat, for his unconditional guidance. He always bestowal parental care upon us and expressd keen interest in solving our problems. An erudite teacher, a tremendous person and a strict disciplinarian, we consider ourselves fortunate to have worked under his supervision. Without his co-operation, the extensive work involved in compiling background information and preparing the paper for publication would not be possible.

## References


[1] Introduction to Biomedical Equipment Technology by Joseph J. Carr and John M. Brown.
[2] Handbook of Biomedical Instrumentation by R. S. Khandpur
[3] Introduction to Medical Electronics Applications by D. Jennings, A Flint, BCH Turton, LDM Nokes
[4] seminarprojects.com/s/biomedical-projects-using-labview
[5] sites.google.com/site/fionaproj/home/beng-401/ecg-circuit
[6] engineerslabs.com/ecg-circuit-analysis-and-design-simulation
[7] www.edaboard.com› Forum › AnalogDesign › Analog Circuit Design
[8] www.docircuits.com/learn/category/ecg/
[9] http://e2e.ti.com/support/amplifiers/precision_amplifiers/f/14/t/148599.aspx
[10] http://www.biosemi.com/publications/artikel5.htm
[11] http://healthcare.analog.com/en/patientmonitoring/ecg-diagnostic-line-powered/segment/health.html
[12] https://www.mathworks.in/products/daq/supported/sound-cards.html
[13] http://www.ni.com/white-paper/5593/en/
[14] www.ni.com/pdf/academic/us/journals/ijee_11.pdf


## Authors

**Bhavikkumar Patel** received B.E.degree in Instrumentation and Control Engineering from the Gujarat Technological University, India. His area of interest covers Bio-medical Instrumentation and Automation aswell as PLC and SCADA.

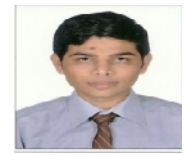

**Dhrumil Shah** has received B.E. degree in Instrumentation and Control Engineering from Gujarat Technological University, India. He also seeks some of his interests in control system designing and Signal Conditioning.

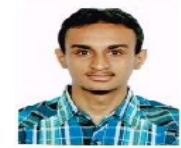